\pdfoutput=1

\documentclass[article,pre,10pt,twocolumn,longbibliography]{revtex4-1}
\usepackage{microtype}
\usepackage{graphicx}
\usepackage[colorlinks,linkcolor=blue,urlcolor=blue,citecolor=blue]{hyperref}

\usepackage{caption}
\usepackage{subcaption}
\usepackage{mathptmx}
\usepackage{times}
\usepackage{epsfig,amsopn}
\usepackage{graphicx}
\usepackage{bm,amsmath,amssymb}
\usepackage{natbib}
\usepackage{braket}
\usepackage{float}
\usepackage{enumerate}
\usepackage{hyperref}
\usepackage{comment}
\allowdisplaybreaks[1]

\begin{document}
\def\be{\begin{equation}}
\def\ee{\end{equation}}
\def\bea{\begin{eqnarray}}
\def\eea{\end{eqnarray}}
\def\f{\frac}
\def\l{\label}
\def\nn{\nonumber}

\definecolor{dgreen}{rgb}{0,0.7,0}
\def\redw#1{{\color{red} #1}}
\def\green#1{{\color{dgreen} #1}}
\def\blue#1{{\color{blue} #1}}
\def\brown#1{{\color{brown} #1}}

\newcommand{\eref}[1]{Eq.~(\ref{#1})}%
\newcommand{\Eref}[1]{Equation~(\ref{#1})}%
\newcommand{\fref}[1]{Fig.~\ref{#1}} %
\newcommand{\Fref}[1]{Figure~\ref{#1}}%
\newcommand{\sref}[1]{Sec.~\ref{#1}}%
\newcommand{\Sref}[1]{Section~\ref{#1}}%
\newcommand{\aref}[1]{Appendix~\ref{#1}}%
\newcommand{\sgn}[1]{\mathrm{sgn}({#1})}%
\newcommand{\erfc}{\mathrm{erfc}}%
\newcommand{\Erf}{\mathrm{erf}}%

\title{First passage under stochastic resetting in an interval}

\author{{\normalsize{}Arnab Pal$^{1,2,3}$}
{\normalsize{}}}
\email{arnabpal@mail.tau.ac.il}

\author{{\normalsize{}V. V. Prasad$^{4}$}
{\normalsize{}}}
\email{prasad.vv@weizmann.ac.il}

\affiliation{\noindent \textit{$^{1}$School of Chemistry, Raymond and Beverly Sackler Faculty of Exact Sciences, Tel Aviv University, Tel Aviv 6997801, Israel}}

\affiliation{\noindent \textit{$^{2}$Center for the Physics and Chemistry of Living Systems. Tel Aviv University, 6997801, Tel Aviv, Israel}}

\affiliation{\noindent \textit{$^{3}$The Sackler Center for Computational Molecular and Materials Science, Tel Aviv University, 6997801, Tel Aviv, Israel}}

\affiliation{\noindent \textit{$^{4}$Department of Physics of Complex Systems, Weizmann Institute of Science, Rehovot 7610001, Israel}}

\date{\today}

\begin{abstract}
We consider a Brownian particle diffusing in a one dimensional interval with absorbing end points. We study the ramifications when such motion is interrupted and restarted
from the same initial configuration. We provide a comprehensive study of the first passage properties of this trapping phenomena. We compute the mean first passage time and derive the criterion on which restart always expedites the underlying completion. We show how this set-up is a manifestation of a success-failure problem.
We obtain the success and failure rates and relate them with the splitting probabilities, namely the probability that
the particle will eventually be trapped on either of the boundaries without hitting the other one. Numerical studies are presented to support our analytic results.

\end{abstract}


\maketitle

\section{Introduction}
The paradigm of diffusion with stochastic resetting has paved our way of understanding restarted processes \cite{Restart1,Restart2}. Consider a Brownian particle which is being reset to a preferred configuration with certain rate. This simple yet pivotal model markedly captures the quintessential features of such processes. There are two cornerstones of this phenomena. In the first case, one is interested in the concentration density of the particles performing such stochastic dynamics. In particular, it has been shown that resetting renders non-equilibrium steady states in generic stochastic processes \cite{Restart1, Restart2,KM,Restart3,restart_conc1, restart_conc2,restart_conc3,restart_conc4,restart_conc5,restart_conc6,restart_conc7,restart_conc8, restart_conc9, restart_conc11,restart_conc12,restart_conc15,restart_conc16,restart_conc17,restart_conc18,restart_conc21,tethered,localtimer,Satya-refractory,Satya-RT}. Canonical examples are diffusion in free space \cite{Restart1,Restart2} or in a potential landscape \cite{restart_conc2}. These studies have also been extended to systems where resets are intermittent \cite{restart_conc3} or the waiting times between reset events are governed by a generic time distribution \cite{Restart4,Restart5}. Moreover, dynamics with resetting exhibit many interesting transient features such as relaxation of the density \cite{Restart4,restart_conc5} to the steady state and its transport properties \cite{transport1,transport2}.

Second, restart has emerged as a conceptual framework to study search processes
\cite{Restart-Search1,Restart-Search2,Restart-Search3,Chechkin,restart_conc19,restart_conc20,ReuveniPRL,PalReuveniPRL,Optimization,drift-diffusion,branching}. Consider a simple diffusive searcher looking for a target. It is well known that the mean search time diverges thus making such strategies undesirable \cite{RednerBook,MetzlerBook,Schehr-review,Benichou-review}. On the contrary, stochastic resetting works in advantage by facilitating long moves from arbitrary location to the resetting location therefore cutting short those long trajectories which are detrimental. Thus restart works in favor of fast completion which otherwise would hinder.
This simple observation has led the researchers to study first passage under generic restart mechanisms. Moreover, it has been demonstrated that it is even possible to find an optimal restart rate which can minimize the mean first passage time. 
Remarkably, one discovers various universality classes displayed by any 
restarted processes which are at optimality \cite{ReuveniPRL,PalReuveniPRL,Optimization}. 

The subject of resetting or restart has been in the limelight recently due to its numerous applications in many interdisciplinary fields.  Apart from applications in search processes or animal foraging, restart has been found to be an indispensable part of chemical reactions \cite{Restart-Biophysics1} and randomized computer algorithms \cite{restart-CS1}. Further progress has seen applications of restart mechanisms in biophysics \cite{Restart-Biophysics2, Restart-Biophysics3, Restart-Biophysics4}, stochastic thermodynamics \cite{restart_thermo1,restart_thermo2}, and quantum mechanics \cite{Quantum1,Quantum2}. 

\begin{figure}[t]
\centering
\includegraphics[width=6.5cm]{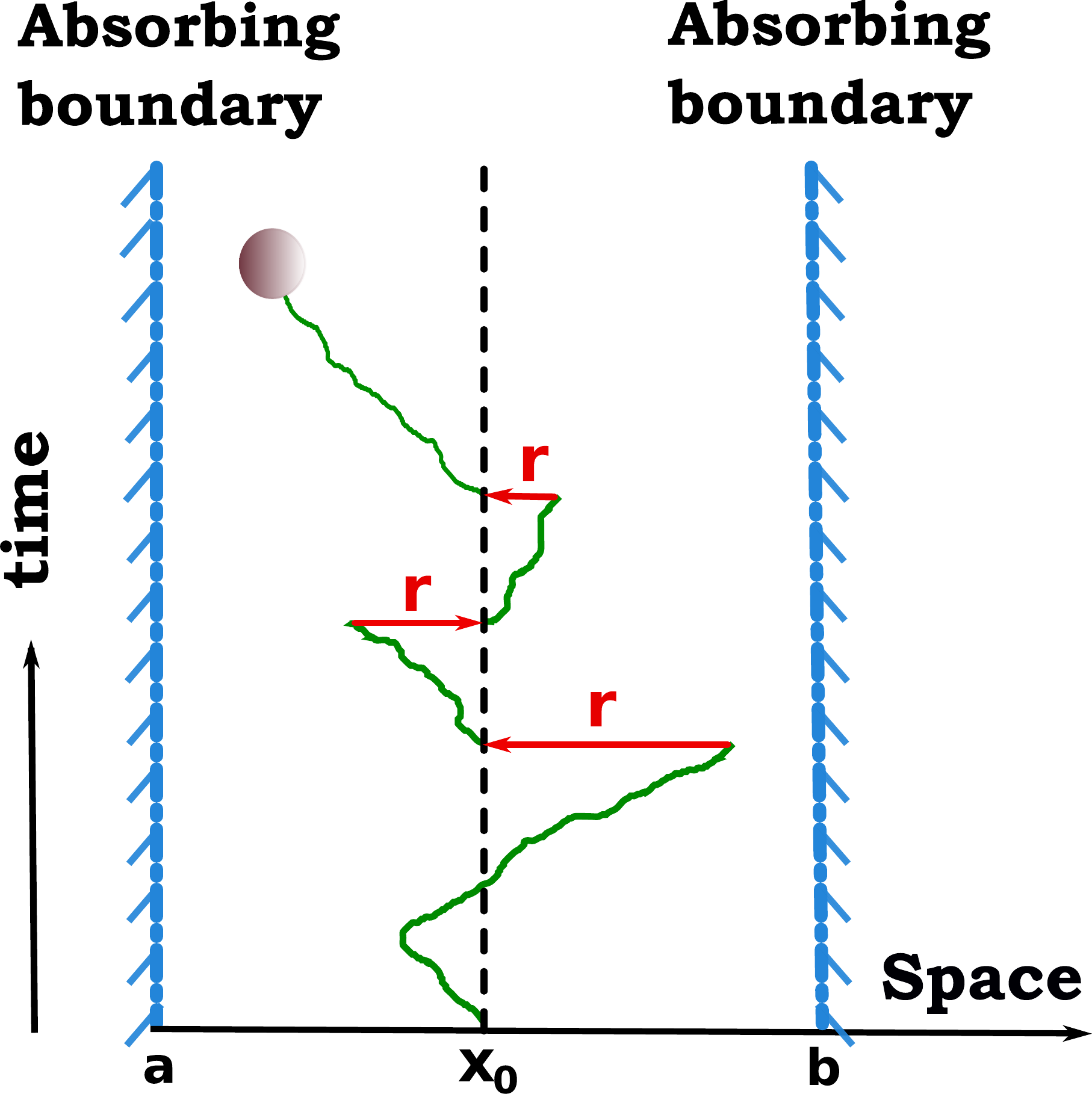}
\caption{(Color online) Schematic of a Brownian particle diffusing in a bounded domain $[a,b]$ in the presence of resetting which reinstates the particle at its initial position $x_0$ with a rate $r$.}
\label{schematic}
\end{figure}

In this paper, we have investigated the motion of a Brownian particle (subjected to resetting) confined in a box $[a,b]$ in one dimension. The boundaries are absorbing (\fref{schematic}). In other words, they can be called targets or possible outcomes. In our set-up, we have two distinct possible outcomes: While the absorption at `$a$' can be defined as a failure or unwanted outcome, absorption at `$b$' can be coined as a success or anticipated outcome. If the underlying process completes prior to reset, then the process immediately ends. Otherwise, restart occurs: the particle is taken back to $x_0$ and allowed to start again given that it was not absorbed meanwhile.
Thus, this model is reminiscent of a first passage process under restart which has different possible outcomes. Another example of such kind is a Bernoulli-like first passage process, which can also end with one of two possibilities.
This set-up  was studied recently in Ref.~\cite{Optimization} within the general framework of first passage under restart \cite{PalReuveniPRL}. In Ref.~\cite{Optimization}, the authors have shown how restart can affect the success or failure probabilities of a Bernoulli trial. Moreover, it was shown that there are optimal restart rates which could maximize or minimize these outcome probabilities. We investigate akin properties by following the motion of the Brownian particle in the box, and
illustrate that by tuning the restart rate it is possible to manipulate  the process to complete in a desired way.
Furthermore, we extend our study to demonstrate how restart can overall accelerate or hinder the completion of the process.
We provide numerical studies to support our results. A similar study on diffusion with resetting in a bounded domain was studied recently in Ref.~\cite{restart_conc8}. However, the boundaries were taken reflecting and an absorbing potential was introduced inside the interval. So, our set-up is very different from that of Ref.~\cite{restart_conc8}.

The paper is organized in the following way. We first compute the survival probability which serves as an essential result of the paper. We illustrate two different methods, namely the backward Fokker-Planck approach (\sref{survival-1}) and the renewal formalism (\sref{survival-2}) to derive this result. This, in turn, allows us to compute the unconditional mean first passage time as a function of restart rate. This we do in \sref{MFPT}. We derive a closed form expression for the unconditional mean first passage time in \sref{MFPT-1}. Using that,
we extract the essential criterion which has to be respected for restart to mitigate the completion in \sref{MFPT-2}. We study the optimal restart rate in \sref{MFPT-3}. \sref{Exit-Time} is devoted to the conditional exit times. We compute the success, failure rates in \sref{Exit-Time-1} and their respective probabilities in \sref{Exit-Time-2}. Furthermore, connections between the unconditional and conditional probabilities are analyzed in details in \sref{Exit-Time-2}. Optimization properties of these probabilities are discussed in \sref{Exit-Time-4}.
Central results of this paper are summarized in \sref{conclusion} with a future outlook. The Appendix contains proofs of some our central results.

\section{Survival Probability}
\label{survival}
We consider a Brownian particle, initially located at $x_0$, diffusing in an interval $[a,b]$ in one dimension. The particle can get absorbed by any of these boundaries. In addition, the particle is stochastically reset to the initial position $x_0$ with a constant rate $r$. We are interested in the first passage properties of the particle to see the trade-off between the resetting and the natural absorption of the particle. To see this, we first provide a through analysis of the survival probability $Q_r(x_0,t)$, defined as the  probability that the particle has hit neither of the boundaries until time $t$ in the presence of resetting, starting from any $x_0$. In other words, it estimates the probability that the particle survives (within the interval) until time $t$. We present two different approaches to derive our results on the survival probability in the presence of resetting.

\subsection{Backward Fokker Planck approach}
\label{survival-1}
It is quite well known that using the backward Fokker Planck
equation can be an advantageous approach to treat the first passage properties. In this case, one first considers the initial position as a variable, and then solve the backward Fokker Planck equation with suitable boundary conditions self-consistently. To this end, we now consider the initial position to be $x$ (which is a variable), while keeping resetting position $x_0$ to be fixed. We solve the equations and at the end set $x$ to be $x_0$.
Following Ref.~\cite{Restart1}, the backward Fokker Planck equation for the survival probability then reads
\bea
\frac{\partial Q_r(x,t)}{\partial t}=D\frac{\partial^2 Q_r(x,t)}{\partial x^2}-r Q_r(x,t)+rQ_r(x_0,t)~,
\label{BFP-Qr}
\eea
where the boundary conditions are $Q_r(a,t)=Q_r(b,t)=0$ and the initial condition is $Q_r(x,0)=1$. The Laplace transform $q_r(x,s)=\int_0^\infty dt e^{-st}Q_r(x,t)$ then satisfies  \cite{Restart1}
\bea
D \frac{\partial^2 q_r(x,s)}{\partial x^2}-(r+s)q_r(x,s)=-1-rq_r(x_0,s)~,
\label{BFP-qr-eq}
\eea
where $q_r(a,s)=q_r(b,s)=0$.
Solving \eref{BFP-qr-eq} with the above boundary conditions, and finally setting
$x=x_0$, we obtain the following expression for the survival probability in the Laplace space
\bea
q_r(x_0,s)=\frac{1-g_r(x_0,s)}{s+rg_r(x_0,s)}~,
\label{BFP-qr}
\eea
where we have defined
\bea
g_r(x_0,s)=\frac{\sinh(b-x_0)\alpha+\sinh(x_0-a)\alpha}{\sinh(b-a)\alpha}~,
\label{gr}
\eea
and $\alpha=\sqrt{\frac{r+s}{D}}$. In the case of one absorbing boundary located at $a=0,$ one recovers $g_r(x_0,s)=e^{-\alpha x_0}$ by setting $b \to \infty$. 
Substituting this expression in \eref{BFP-qr}, we find $q_r(x_0,s)=\frac{1-e^{-\alpha x_0}}{s+re^{-\alpha x_0}}$, as derived earlier in Ref.~\cite{Restart1}.

\subsection{Renewal approach}
\label{survival-2}
One can also realize the restarted processes within the elegant formalism of renewals where one makes use of the fact that upon each reset, the system renews itself. Following Ref.~\cite{Restart4}, we can write the survival probability in the following way
\bea
Q_r(x_0,t)=e^{-rt}Q_0(x_0,t)+r\int_0^t d\tau e^{-r\tau} Q_0(x_0,\tau)Q_r(x_0,t-\tau)~, \nonumber \\
\label{Qr-renewal}
\eea
where $Q_0(x_0,t)$ is the survival probability of the particle in the interval upto time $t$ in the absence of resetting. \eref{Qr-renewal} has a simple interpretation. The first term on the right hand side implies that the particle survives until time $t$ without experiencing any reset event. The second term considers the possibility when there are multiple reset events. One can then look at a long trajectory where the last reset event had occurred at time $t-\tau$, and after that there has been no reset for the duration $\tau$. This probability is given by $rd\tau e^{-r\tau}$. But then this has to be multiplied by $Q_r(x_0,t-\tau)$, i.e., the probability that the particle survives until time $t-\tau$ with multiple reset events and $Q_0(x_0,\tau)$, i.e., the survival probability of the particle for the last non-resetting interval $\tau$ \cite{Restart4}. The Laplace transform $q_r(x_0,s)$ then satisfies
\bea
q_r(x_0,s)=\frac{q_0(x_0,s+r)}{1-rq_0(x_0,s+r)}~,
\label{renewal-qr}
\eea
where $q_0(x_0,s)$ is the Laplace transform of $Q_0(x_0,t)$. Following Refs.~\cite{RednerBook, MetzlerBook, Schehr-review}, we use the well-known expression for  $Q_0(x_0,t)$
\bea
Q_0(x_0,t)=2\sum_{n=1}^{\infty}\psi_n(x_0)\phi(n)~e^{-k_n t}~,
\label{Q0}
\eea
where $\psi_n(x)=\sin\bigg[\dfrac{(x-a)n\pi}{b-a}\bigg]$ are the eigenfunctions with $\phi(n)=\frac{1-\cos(n\pi)}{n\pi}$. Also $k_n=n^2 \pi^2D/(b-a)^2$ is the rate at which the $n$-th eigenmode $\psi_n(x)$ decays with time. Thus the longest decay time $k_1^{-1}=(b-a)^2/D\pi^2$ characterizes the diffusing dynamics within the interval.
Making use of the renewal formula [\eref{renewal-qr}], we obtain an expression of the Laplace transform $q_r(x_0,s)$,
\bea
q_r(x_0,s)=\frac{2 \sum_{n=1}^{\infty}\psi_n(x_0)\phi(n)/\Delta(n,r,s)}{1-2r\sum_{n=1}^{\infty}\psi_n(x_0)\phi(n)/\Delta(n,r,s)}~,
\label{renewal-qr-LT}
\eea
where 
\bea
\Delta(n,r,s)=k_n+r+s~.
\label{delta}
\eea
The expressions obtained via backward Fokker-Planck equation (\eref{BFP-qr}) or the renewal approach (\eref{renewal-qr-LT}) are equivalent (see \aref{equivalence} for more details). 

\subsection{Position density of the particle}
\label{prob}
In the presence of two absorbing boundaries, the particle will be absorbed in the absence of resetting. Question is by relocating the particle repeatedly to its initial position whether one can still find a finite probability to observe the particle within the domain at large time. To see this, we first define the position density $P_r(x,t|x_0,0)$, which estimates the probability of finding the particle at $x$ at time $t$ given that it had started from $x_0$ in the presence of multiple resetting at $x_0$. We can write a time dependent equation for the position density $P_r(x,t|x_0,0)$ using a renewal formalism
\bea
P_r(x,t|x_0,0) &=& e^{-rt}P_0(x,t|x_0,0) \nonumber \\
&+& r\int_0^t~d\tau~e^{-r \tau}~P_0(x,\tau|x_0,0) Q_r(x_0,t-\tau)~,
\label{propagator-renewal}
\eea
where $P_0(x,t|x_0,0)$ is the probability density of finding the particle in the interval in the absence of resetting. Also, recall that $Q_r(x_0,t)=\int_a^b~dx~P_r(x,t|x_0,0)$ is the survival probability until time $t$. We emphasize that \eref{propagator-renewal} can be interpreted in an identical manner as in \eref{Qr-renewal}. By taking Laplace transform on both sides of \eref{propagator-renewal} and using
\eref{renewal-qr}, we find
\bea
p_r(x,s|x_0,0)=\frac{p_0(x,s+r|x_0,0)}{1-r q_0(x_0,s+r)}~.
\label{propagator-formula}
\eea
The probability density $P_0(x,t|x_0,0)$ is a classical result and known from the literature \cite{RednerBook}
\bea
P_0(x,t|x_0,0)&=&\dfrac{2}{b-a}\sum_{n=1}^{\infty}  \psi_n(x_0)\psi_n(x) e^{-k_n t}~.
\label{propagator-no-reset}
\eea
Using the Laplace transform of $P_0(x,t|x_0,0)$ and substituting $q_0(x_0,s)$ in \eref{propagator-formula} one obtains
\bea
&p_r&(x,s|x_0,0)=\frac{\alpha}{2} \nonumber \\
&\times& \frac{\cosh \left[ (b-a-|x-x_0|)\alpha  \right]-\cosh \left[ (b+a-x_0-x)\alpha  \right]}{s~ \sinh \left[ (b-a)\alpha  \right]+r\sinh \left[ (x_0-a)\alpha  \right]+r\sinh \left[ (b-x_0)\alpha  \right]}
\label{propagator-reset}
\eea
In order to find the asymptotic behavior of the position density (i.e., the steady state), we can make use of the final value theorem. The theorem asserts that $P_{ss}(x)=\lim_{t \to \infty} P_r(x,t|x_0,0)=\lim_{s \to 0}s p_r(x,s|x_0,0)$ \cite{Restart4}. By doing a $s\to 0$ expansion on the right hand side of \eref{propagator-reset}, we note that there are no terms which are of order $1/s$, for any resetting rate $r$. 
This indicates an absence of finite value for the probability density 
in the large time limit i.e., no particle will survive given long enough time. Once the spread of the concentration becomes comparable to the interval size, the flux of probability through the boundaries becomes significant and the density inside the interval decays rapidly despite the resetting dynamics. 

\section{Unconditional First Passage Time}
\label{MFPT}

In the preceding section, we have studied the survival probability of the Brownian particle in the interval bounded between $a$ and $b$. It is only natural to ask the time it will take for the particle to exit or get absorbed for the first time by any of these boundaries in the presence of resetting at $x_0$. These first passage times are random, and one is in general interested in the statistics of such time. In this section, we will focus on this time statistics in the presence of resetting. It is worth mentioning that this situation is analogous to a first passage process with multiple outcomes, and the cumulative first passage time statistics for any possible outcome is statistically identical to that of unconditional first passage time.

\subsection{Mean first passage time}
\label{MFPT-1}
\begin{figure}[t]
\centering
\includegraphics[width=8cm]{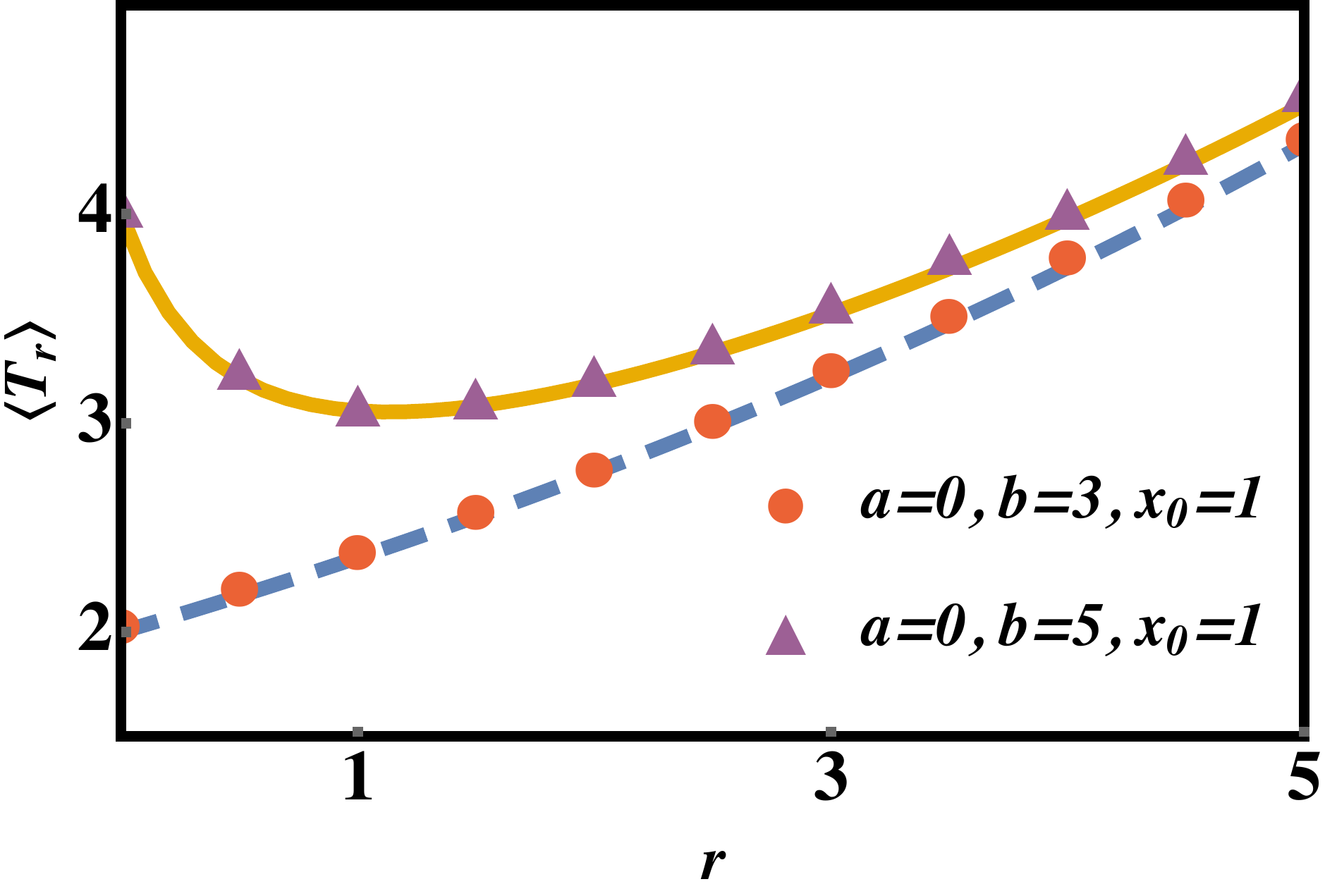}
\caption{(Color online) Plot for the unconditional mean first passage time as a function of restart rate for two different boundary conditions: (i)~$a=0,~b=3$, and (ii)~$a=0,~b=5$, with the initial condition fixed at $x_0=1$ for both cases. The diffusion constant has the value $D=1/2$. The theoretical formulas [dashed and solid lines for (i) and (ii) respectively], as in \eref{Tr-2}, are corroborated by their respective simulation data (in markers).}
\label{fig MFT two cases}
\end{figure}

We will first characterize the average time it takes for a particle to exit through any of the boundaries. We call this unconditional mean first passage time, denoted by $\langle T_r(x_0) \rangle$. By noting that $\langle T_r(x_0) \rangle=\int_0^\infty dt~ t f_{T_r}(t)$, where $f_{T_r}(t)=-\frac{\partial Q_r(x_0,t)}{\partial t}$, is the unconditional first passage time density. Hence, by doing integration by parts, one can write
\bea
\langle T_r(x_0) \rangle=q_r(x_0,s \to 0)= \frac{1}{r} \left[ \frac{1}{g_r(x_0,0)}-1 \right]~,
\label{Tr-1}
\eea
where we have used \eref{BFP-qr}. Further simplification using \eref{gr} leads us to the following expression for the mean first passage time
\bea
\langle T_r(x_0) \rangle=\frac{1}{r} \left[ \frac{\sinh(b-a)\alpha_0}{\sinh(b-x_0)\alpha_0+\sinh(x_0-a)\alpha_0}-1  \right]~,
\label{Tr-2}
\eea
where
\bea
\alpha_0=\alpha|_{s\to 0}=\sqrt{\frac{r}{D}}~.
\label{alpha0}
\eea
In the limit of vanishing restart rate ($r \to 0$), we obtain $\langle T_0 \rangle=(x_0-a)(b-x_0)/2D$ \cite{RednerBook}.
We can also obtain $\langle T_r(x_0) \rangle$ using the renewal formula obtained in Ref.~\cite{PalReuveniPRL}, which requires a prior knowledge of the underlying first passage time distribution (without resetting). 
In \fref{fig MFT two cases}, we have plotted $\langle T_r \rangle$ using \eref{Tr-2} as a function reset rate for two different set of parameters: (i) $a=0,~b=3$, and (ii) $a=0,~b=5$ for fixed $x_0=1$. In the first case $\langle T_r \rangle$ increases monotonically with rate $r$, unlike in the latter case. In fact, in the second case, restart lowers the mean first passage time and thus accelerating the completion. Nevertheless, it is not evident what sets the criterion for restart either to prolong or expedite the completion. To characterize this transition, we do a detailed analysis of the restart criterion in the next subsection.

\subsection{Analysis of the restart criterion} 
\label{MFPT-2}
It is well understood from the theory of first passage under restart that restart has the ability to expedite the underlying completion if $\text{CV}>1$, where $\text{CV}$ stands for the ratio between the standard deviation $\sigma(T_0)$ and the mean first passage time $\langle T_0 \rangle$ of the underlying (i.e., without restart) first passage time process. One can arrive at this criterion by simply setting
$\frac{d \langle T_r \rangle}{dr}|_{r\to 0}<0$ \cite{PalReuveniPRL}.
In the case of a diffusive first passage process (where the first passage time distribution is given by L\'evy-Smirnov distribution \cite{RednerBook, MetzlerBook, Schehr-review}), the criterion is inherently satisfied. 
Thus restart will always expedite the completion of such process. On the other hand, as we have shown in \fref{fig MFT two cases}, restart can both expedite or hinder the completion of a diffusive motion in an interval. To understand the criterion in terms of the relevant parameters of the system, we first set the criterion $\text{CV}>1$ which yields
\bea
L^2+3L+1-5(L+1)u+5u^2>0~,
\label{criterion}
\eea
where $L=a/b$ and $u=x_0/b$. We arrive at this expression by recalling that the mean first passage time is given by $\langle T_0 \rangle=(x_0-a)(b-x_0)/2D$ and the second moment is given by $\langle T_0^2 \rangle=-2\left[ dq_0(x_0,s)/ds \right]|_{s \to 0}$.
Obtaining $q_0(x_0,s)$ from \eref{BFP-qr} by taking $r\to 0$ limit yields $\langle T_0^2 \rangle=(x_0-a)(b-x_0)(a^2-3ab+b^2+ax_0+bx_0-x_0^2)/12D^2$.
Substituting these expressions in the criterion $\text{CV}>1$, we obtain \eref{criterion}. A crucial observation is that the criterion is independent of the diffusion coefficient $D$ (hence, the motion of the particle), thus only the combination of length scales $(x_0,a,b)$ will set the criterion. Further simplifications can be made if we choose $a=0$. Then from \eref{criterion}, we obtain $5u^2-5u+1>0$. This determines the domain in which restart expedites the completion of the underlying process: $\mathcal{D}=[ (0, u_{-}) ~\cup~ (u_{+},1) ]$, where $u_{\pm}=(5\pm \sqrt{5})/10$. Hence, if the particle starts closer to either of the boundaries (i.e. $0<x_0<bu_{-}$, or $bu_{+}<x_0<b$), the completion will be accelerated. 
 On the other hand, if $u_{-}<u<u_{+}$, restart will not be beneficial. This means if the particle starts in the region $bu_{-}<x_0<bu_{+}$ which is centered around $x_0=b/2$, restart will only prolong the completion.

 \subsection{Optimal restart rate}
 \label{MFPT-3}
\begin{figure}[h]
\centering
\includegraphics[width=8.cm]{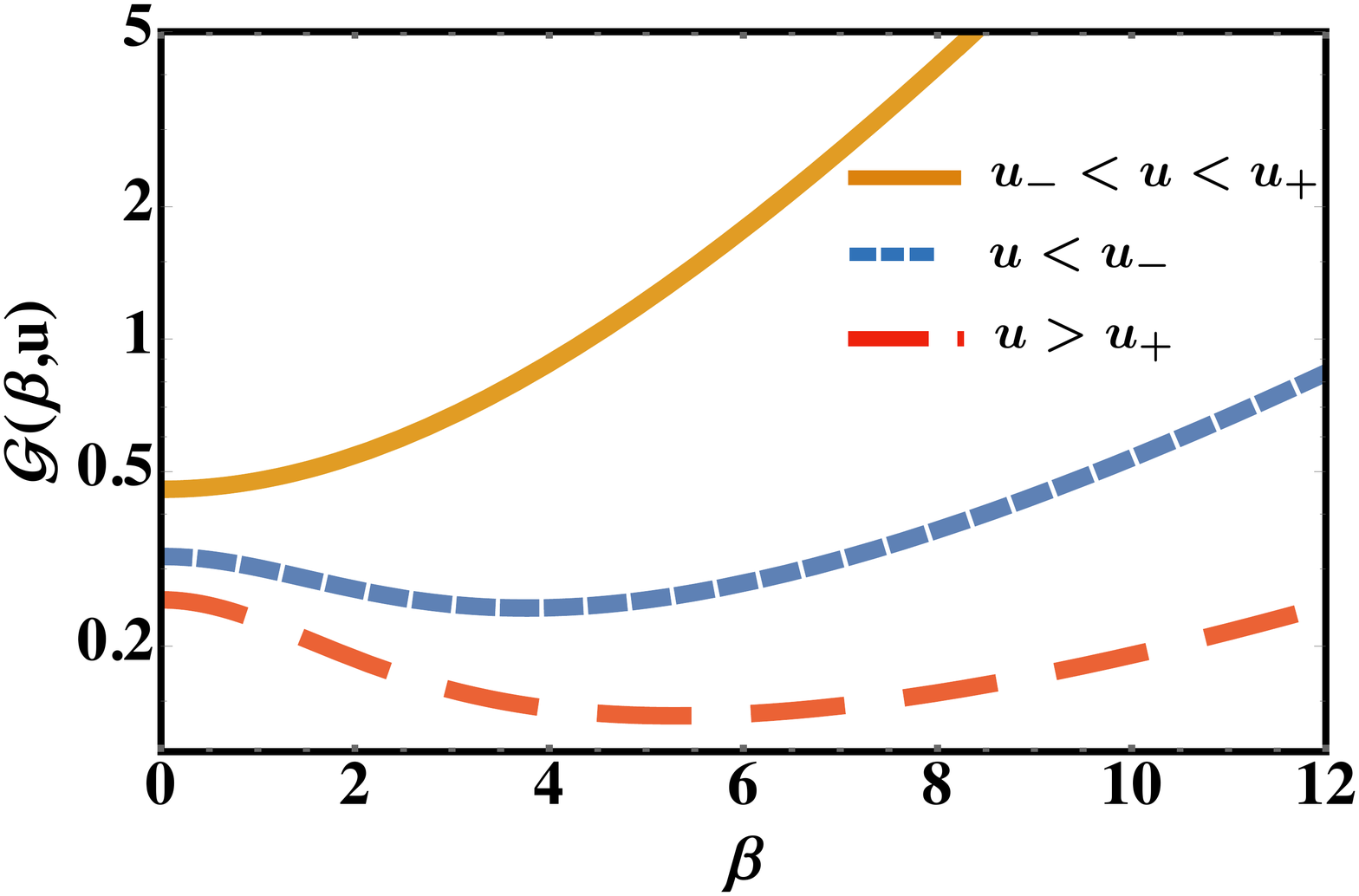}
\caption{(Color online) Plot for the scaling function $\mathcal{G}(\beta,u)$~[\eref{scaling-function}] as a function of $\beta$. The function has minima when $u$ takes value from the domain $\mathcal{D}$, as shown for $u=0.2<u_{-}$ (short dashed line in the middle), and $u=0.85>u_{+}$ (long dashed line at the bottom). 
However, when $u_{-}<u=0.65<u_{+}$ (solid line at the top),
there is no minimum for $\mathcal{G}(\beta,u)$ at any finite $r$, indicating that restart is not beneficial.
}
\label{optimal}
\end{figure}

 The above analysis asserts that restart can accelerate the completion of the underlying process in certain regimes of the parameter values. This regime was fully characterized in the previous section and also will be a focus of our analysis here. In this regime, restart not only expedites the completion, one also observes that the mean first passage time can be minimized at a certain value of restart rate. To further illustrate on this optimal restart rate, we first note that \eref{Tr-2} can be scaled in the following way
 \bea
 \langle  T_r \rangle=\frac{b^2}{4D}\mathcal{G}(\beta,u)~,
 \label{scaling}
 \eea
after setting the left boundary $a=0$. In \eref{scaling}, we have defined a new scaling variable $\beta=\frac{b}{2}\alpha_0$ and the scaling function $\mathcal{G}(\beta,u)$, which is now a function of restart rate for a given set of $(x_0,b)$, is given by
\bea
\mathcal{G}(\beta,u)=\frac{1}{\beta^2}\left[\frac{\cosh(\beta)}{\cosh \beta(1-2u)}-1\right]~.
\label{scaling-function}
\eea
To find the optimal restart rate, one sets $\frac{\partial \mathcal{G}}{\partial \beta}|_{\beta=\beta^*}=0$. This, in turn, gives the optimal restart rate $r^*=4D{\beta^*}^2/b^2$, in terms of $\beta^*$. In \fref{optimal}, we optimize the scaling function $\mathcal{G}(\beta,u)$ numerically as a function of $\beta$ in three different regimes of $u$ defined by the domain $\mathcal{D}$. It is evident from \fref{optimal} that when $u_{-}<u<u_{+}$, the function is minimum at $\beta^*=0$, meaning $\langle T_r \rangle$ can not be made lower by introducing a finite restart rate. On the other hand, when $u \in \mathcal{D}$, we see that the scaling function $\mathcal{G}(\beta,u)$ is minimized at a finite $\beta$ implying that the mean first passage time is optimized at a finite restart rate.

\section{Conditional First Passage Time}
\label{Exit-Time}

\begin{figure}[b]
\centering
\includegraphics[width=6.05cm]{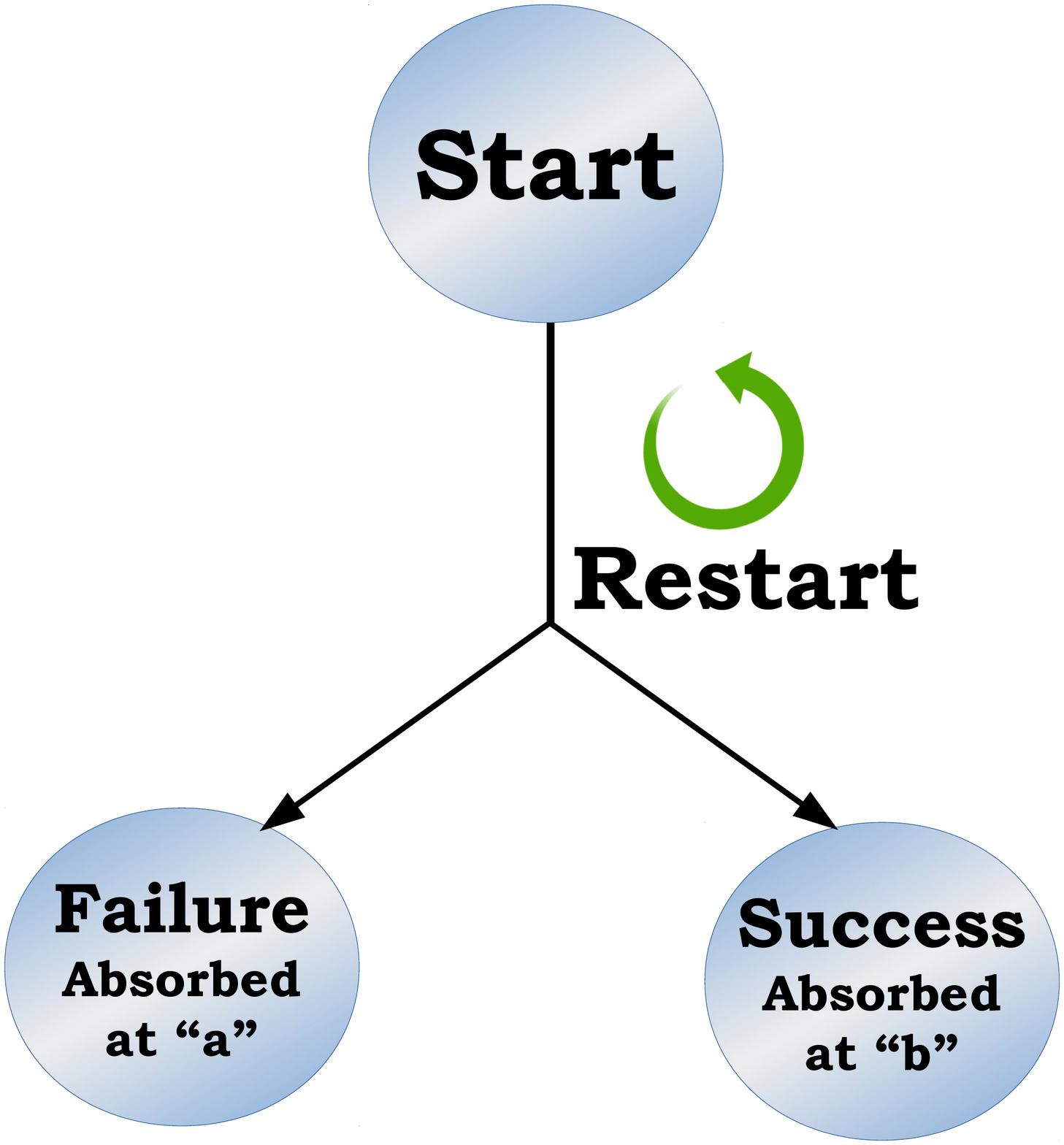}
\caption{(Color online) Depiction of first passage under resetting in an interval as a success-failure problem. A `successful' event is defined by an absorption at the boundary `$b$', while a `failure' event occurs when an absorption takes place at the boundary `$a$'.}
\label{success-failure}
\end{figure}

So far, we have focused only on the unconditional mean first passage time i.e., the exit time irrespective of boundaries or the particular choice of outcomes within the success-failure set-up (see \fref{success-failure}).
We devote this section to study observables conditioned on the `outcome of our choice'. Examples of such observables are: conditional mean exit times from boundary `$b$' or `$a$'. Recalling that a `successful' event was defined by an absorption at the boundary `b' and `failure' by absorption at the boundary `a', the conditional exit times provide us estimations of success or failure \textit{rates}. Another natural quantity to investigate would be the splitting probability namely the probability to escape through a specified boundary without hitting the other ones which is, in fact, a measure of success or failure \textit{probability}. Interestingly, restart could optimize the success or failure probabilities. In other words, by modulating restart rate one can reduce the occurrence of failure events, while facilitating the probability of success to its maximum. 

\subsection{Mean conditional exit times}
\label{Exit-Time-1}

\begin{figure}[t]
\centering
\includegraphics[width=8cm]{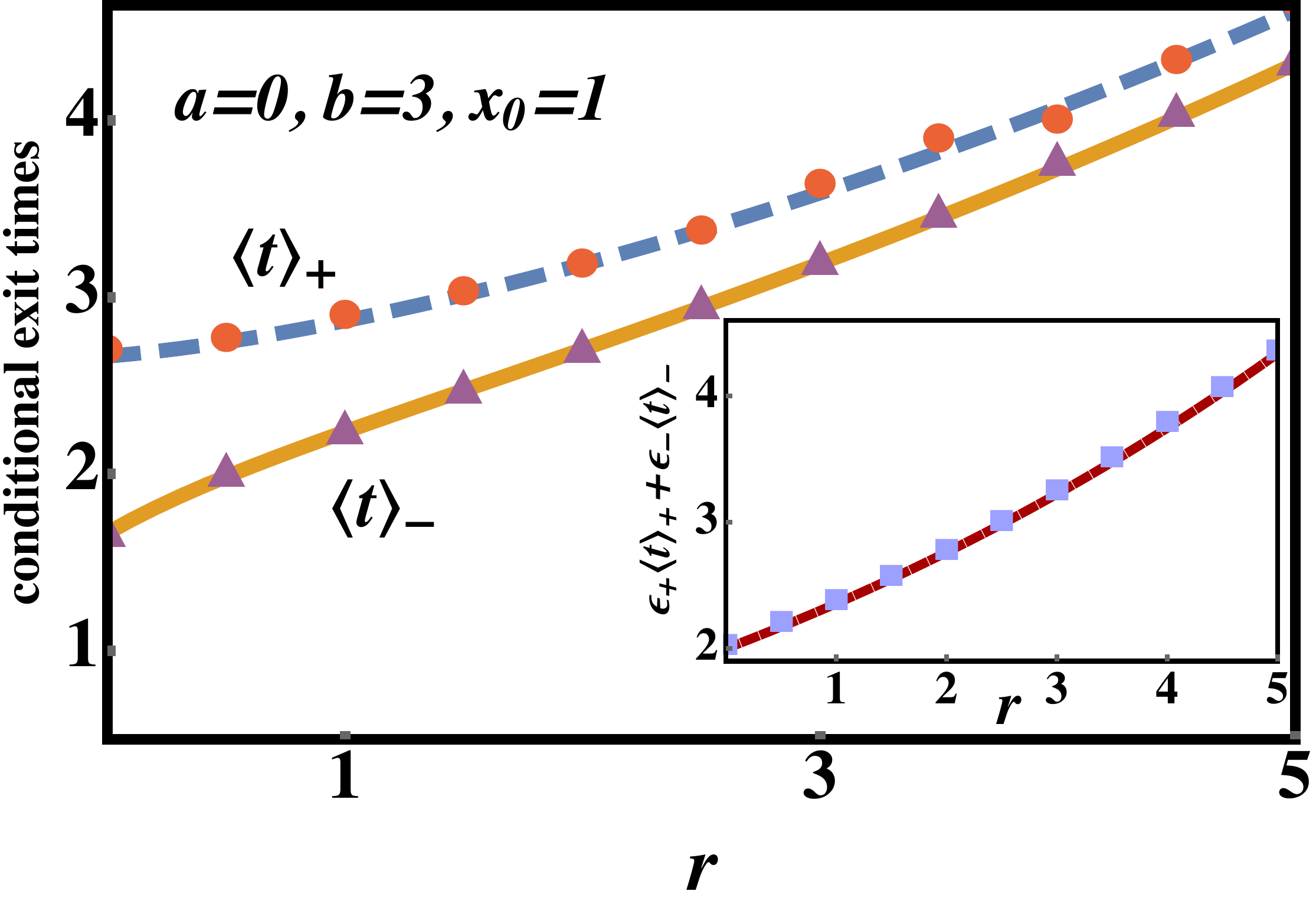}
\caption{(Color online) We have plotted the conditional exit times $\langle t \rangle_{+}$ and $\langle t \rangle_{-}$ as a function of restart rate $r$ for a system  with the end points ~$a=0,~b=3$ and resetting position $x_0=1$ with $D=1/2$. The dashed and solid lines represent analytical expressions for $\langle t \rangle_{+}$ and $\langle t \rangle_{-}$ (given by \eref{conditional-exit-time}) respectively. Markers in circle and triangle represent the corresponding simulation data points. The inset shows the plot for the function $\epsilon_{+}\langle t \rangle_{+}+\epsilon_{-}\langle t \rangle_{-}$~(dashed line) against $\langle T_r \rangle$ (markers) as a function of restart rate. The match between the two provides a numerical demonstration of \eref{relation-MFT-conditional exit times}.}
\label{fig conditional time}
\end{figure}

Mean conditional exit time is the average time for the particle to hit a specific boundary without hitting the other one. Let us define $\langle t(x_0) \rangle_{\pm}$ to be the conditional mean exit times through the boundaries $b$ and $a$ respectively. 
To compute this time statistics, it is best to first measure the flux of current flowing through each one of these boundaries. Probability flux current at any point $x$ in space at time $t$ is defined by $J(x,t)=-D\frac{\partial P_r(x,t|x_0,0)}{\partial x}$. The exit times and the flux are then related to each other by the following relation \cite{RednerBook}
\bea
\langle t(x_0) \rangle_{\pm}=\frac{\int_0^\infty dt~t~J_{\pm}(x_0,t)}{\int_0^\infty dt~J_{\pm}(x_0,t)}~,
\label{mean-exit-times-definition}
\eea
where $J_{\pm}(x_0,t)$ denote the currents at each of the boundaries, as given below:
\bea
J_{+}(x_0,t)&=&-D\frac{\partial P_r(x,t|x_0,0)}{\partial x}|_{x=b}~, \nonumber \\ J_{-}(x_0,t)&=&D\frac{\partial P_r(x,t|x_0,0)}{\partial x}|_{x=a}~.
\label{conditional-currents}
\eea
The expression for the conditional exit times \eref{mean-exit-times-definition} can be understood as follows.
Note that the currents, defined in \eref{conditional-currents}, are identical to the conditional first passage time densities through each one of these boundaries. Hence, to obtain $\langle t(x_0) \rangle_{\pm}$, one needs to average over these conditional first passage time densities. 
The conditional mean exit times in \eref{mean-exit-times-definition} can be rewritten in terms of the currents in Laplace space
\bea
\langle t(x_0) \rangle_{\pm}=\frac{-\frac{\partial j_{\pm}(x_0,s)}{\partial s}|_{s \to 0}}{j_{\pm}(x_0,s=0)}~,
\label{mean-exit-times-definition-LT}
\eea
where $j_{\pm}(x_0,s)=\int_0^{\infty}dt~e^{-st}~J_{\pm}(x_0,t)$, satisfy
\bea
j_{+}(x_0,s)&=&-D\frac{\partial p_r(x,s|x_0,0)}{\partial x}|_{x=b}~, \nonumber \\ j_{-}(x_0,s)&=&D\frac{\partial p_r(x,s|x_0,0)}{\partial x}|_{x=a}~.
\label{conditional-currents-LT}
\eea
Substituting $p_r(x,s|x_0,0)$ in \eref{conditional-currents-LT} from \eref{propagator-reset}, we get the following expressions for the currents
\bea
j_{+}(x_0,s)&=&
D\alpha^2 \sinh\left[(x_0-a)\alpha  \right]/\mathcal{F}(s,r), \nonumber \\
j_{-}(x_0,s)&=&
D\alpha^2 \sinh\left[(b-x_0)\alpha  \right]/\mathcal{F}(s,r)~,
\label{conditional-currents-LT-exact}
\eea
where 
\bea
\mathcal{F}(s,r)=&&s\sinh \left[ (b-a)\alpha  \right]+\nonumber\\&& r \left(\sinh \left[ (x_0-a)\alpha  \right]+ \sinh \left[(b-x_0)\alpha  \right]\right) 
\label{mathcalF}
\eea
Now using \eref{conditional-currents-LT-exact} and \eref{mathcalF} in \eref{mean-exit-times-definition-LT} one gets the exact formula for the conditional exit times 
\begin{equation}
\begin{split}
\langle t(x_0) \rangle_{+}&=\frac{ \mathcal{F}_1(x_0-a,b-x_0,b-a)
}{2D\alpha_0^2(1+\text{cosech}(\alpha_0[x_0-a])\sinh(\alpha_0[b-x_0]))~}~,\\\\
\langle t(x_0) \rangle_{-}&=\frac{\mathcal{F}_1(b-x_0,x_0-a,b-a)}{2D\alpha_0^2(1+\text{cosech}(\alpha_0[b-x_0])\sinh(\alpha_0[x_0-a]))~}~
\end{split}
\label{conditional-exit-time}
\end{equation}
where
\bea
&\mathcal{F}_1&(k_1,k_2,k_3)=-2+ \text{cosech}(\alpha_0k_1)\bigg[\alpha_0k_2\cosh(\alpha_0k_2)\nonumber\\&+&2\sinh(\alpha_0k_3)+\Big(-2+\alpha_0k_1\coth(\alpha_0k_1)\Big)\sinh(\alpha_0k_2)\bigg]\nonumber\\
\eea
\noindent
The success and failure rates are then given by
\bea
k_s=\langle t(x_0) \rangle_{+}^{-1},~~~k_f=\langle t(x_0) \rangle_{-}^{-1}~.
\eea
Note that the conditional exit times are monotonic function of the restart rate, and moreover they diverge as restart rate increases (see \fref{fig conditional time}). In the limit of vanishing restart rate, the mean conditional exit times through the boundaries $b$ and $a$ are given by
\bea
\langle t(x_0) \rangle_{+}|_{r\to 0}&=&\frac{1}{6D}(b-x_0)(b+x_0-2a)~, \nonumber \\
\langle t(x_0) \rangle_{-}|_{r\to 0}&=&\frac{1}{6D}(x_0-a)(2b-x_0-a)~.
\label{conditional-exit-time-req0}
\eea
In \fref{fig conditional time}, we have 
plotted simulation data against the theoretical formulas (obtained in \eref{conditional-exit-time}) for conditional exit times as a function of restart rate, for a system with boundaries $a=0,~b=3$ and the resetting position $x_0=1$, when $D=1/2$. The $r \to 0$ limit values are also in accordance with \eref{conditional-exit-time-req0}, as expected.

\subsection{Splitting probabilities}
\label{Exit-Time-2}
We present now analytical expressions for the splitting probabilities, the probability that the process starting at $x_0$ and evolving in a box $[a, b]$ hits the right boundary $b$ before hitting the left boundary at $a$, and vice-versa. Let us define these probabilities as $\epsilon_{\pm}(x_0)$ respectively. Thus, $\epsilon_{+}, \epsilon_{-}$ denote the success and failure probability respectively.
To compute these probabilities, one needs to integrate the current flux flowing through each of the boundaries over the time as shown below
\bea
\epsilon_{\pm}(x_0)=\int_0^\infty~dt~J_{\pm}(x_0,t)=j_{\pm}(x_0,s=0)~.
\label{epsilon}
\eea
Using the formulae for $j_{\pm}(x_0,s)$ from \eref{conditional-currents-LT-exact} and 
substituting for $s=0$, we obtain the following expressions for the splitting probabilities:
\bea
\epsilon_{+}(x_0)&=&\frac{\sinh \left[ (x_0-a)\alpha_0  \right]}{\sinh \left[ (x_0-a)\alpha_0  \right]+\sinh \left[ (b-x_0)\alpha_0  \right]}, \\
\epsilon_{-}(x_0)&=&\frac{\sinh \left[ (b-x_0)\alpha_0  \right]}{\sinh \left[ (x_0-a)\alpha_0  \right]+\sinh \left[ (b-x_0)\alpha_0  \right]}.
\eea

\begin{figure}[b]
\centering
\includegraphics[width=9.cm]{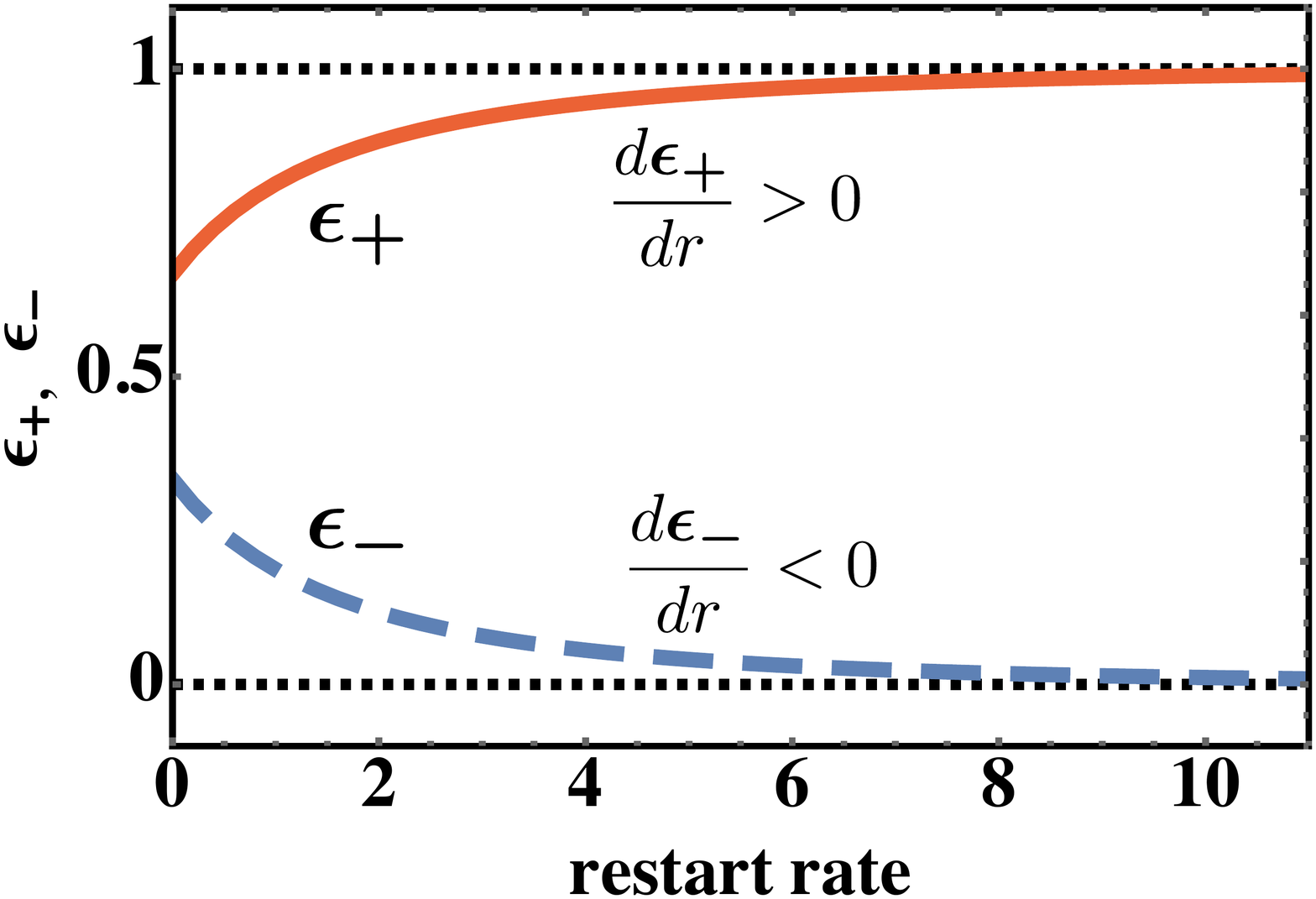}
\caption{(Color online) Plot of splitting probabilities $\epsilon_{+}(x_0)$ and $\epsilon_{-}(x_0)$ as a function of restart rate (in solid and dashed lines respectively). The parameters are taken as $a=0,b=3,x_0=2$, and $D =1/2$.}
\label{r-opt}
\end{figure}

At $r\to 0$ limit, these probabilities of winning or losing will be given by $\epsilon_{+}|_{r \to 0}=\frac{x_0-a}{b-a}, \epsilon_{-}|_{r \to 0}=\frac{b-x_0}{b-a}$. Further taking $a\to 0$, we find $\epsilon_{+}=\frac{x_0}{b}, \epsilon_{-}=1-\frac{x_0}{b}$. Thus the
probability of reaching one end point is just the relative distance to the other end point, which is quite remarkable \cite{RednerBook}. On the other hand, in the large $r$ limit, behavior of splitting probabilities has a strong dependence on the resetting (initial) position $x_0$. If $x_0>\frac{a+b}{2}$, we have $\epsilon_{+} \to 1,~ \epsilon_{-} \to 0$ i.e., the particle which starts (and resets) on the positive side of the middle of the interval eventually reaches the boundary $b$ with probability one. Conversely, when $x_0<\frac{a+b}{2}$, one finds $\epsilon_{+} \to 0,~ \epsilon_{-} \to 1$, which is exactly one would expect. However, if $x_0=\frac{a+b}{2}$ then we have $\epsilon_{\pm}=\frac{1}{2}$ for any restart rate.
In \fref{r-opt}, we have plotted the splitting probabilities as a function of restart rate. It is also evident from the figure that these probabilities saturate either to maximum or minimum as a function of restart rate. In \sref{Exit-Time-4}, we will show how one can utilize this particular property to engineer restart as a tool to the fulfillment of the required outcomes.

A careful observation of the exact expressions for the conditional exit times and the splitting probabilities immediately leads us to establish the following relation
\bea
\langle T_r \rangle=\epsilon_{+}\langle t \rangle_{+}+\epsilon_{-}\langle t \rangle_{-}~,
\label{relation-MFT-conditional exit times}
\eea
which holds for any restart rate.
This means that
the unconditional mean exit time (independent
of which side is exited) is the appropriately weighted average of the
conditional mean exit times to each boundary. These weights are the conditional splitting probabilities. But this is no surprise and can be understood by simple path enumeration. The splitting probability $\epsilon_{+}(x_0)$ sums over all the paths that start from $x_0$ and exit through boundary $b$ without hitting the boundary $a$. Thus,  $\epsilon_{+}(x_0)=\sum_{p_{+}}\sigma_{p_{+}}(x_0)$, where $\sigma_{p_{+}}(x_0)$ denotes the weight of a single trajectory from $x_0$ to $b$ that avoids $a$. On the other hand, $\langle t(x_0) \rangle_{+}$ is the exit time through boundary $b$ conditioned on the fact that it has survived the boundary $a$. Thus, $\langle t(x_0) \rangle_{+}=\frac{\sum_{p_{+}}\sigma_{p_{+}}(x_0)t_{p_{+}}(x_0)}{\sum_{p_{+}}\sigma_{p_{+}}(x_0)}$, where $t_{p}(x_0)$ is the exit time of a specific trajectory that starts at $x_0$ through a boundary. One can use a similar argument for $\epsilon_{-}(x_0)$ and $\langle t(x_0) \rangle_{-}$. \eref{relation-MFT-conditional exit times} is also demonstrated in the inset of \fref{fig conditional time}.

\begin{figure}[t]
\centering
\includegraphics[width=8.cm]{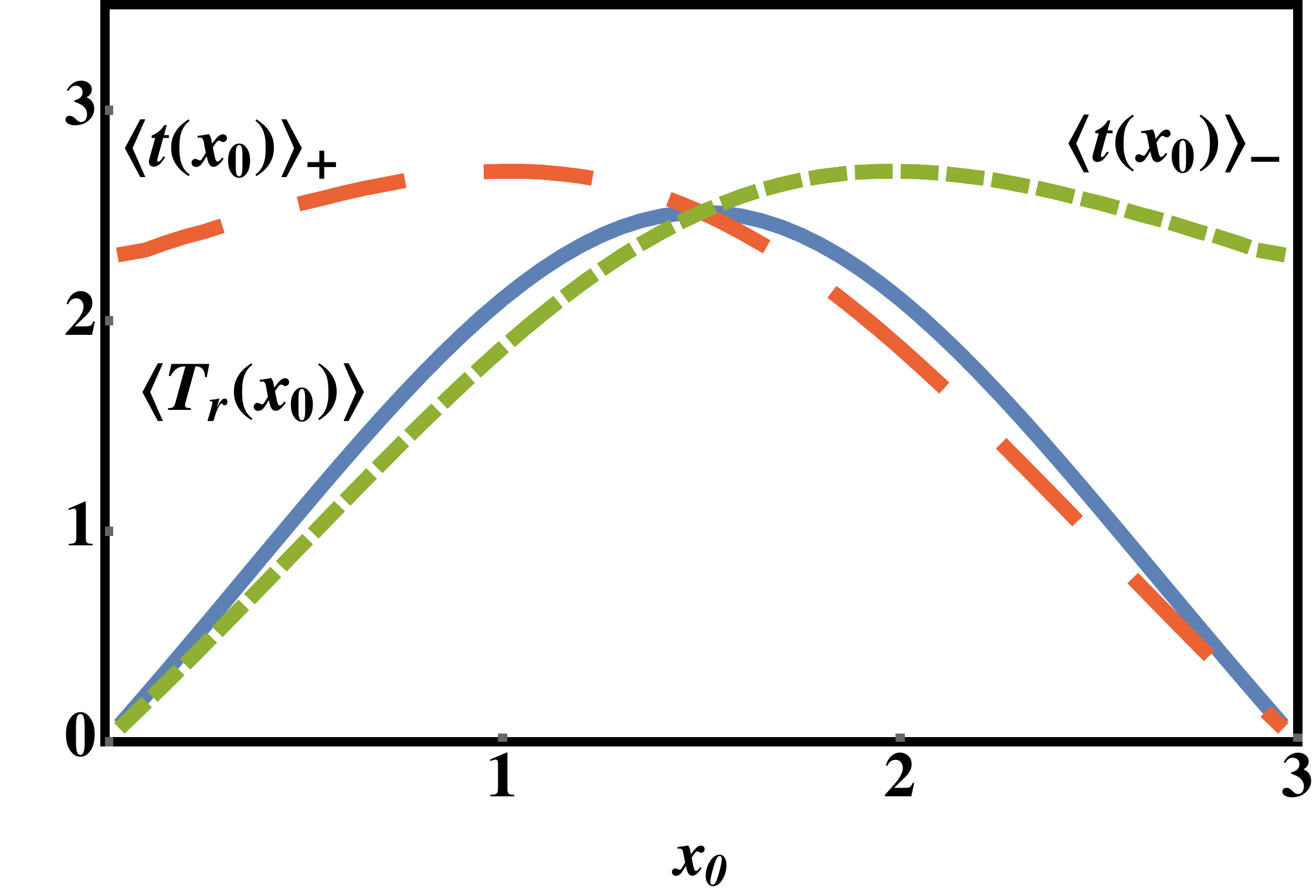}
\caption{(Color online) Plot for the conditional mean exit times $\langle t \rangle_{\pm}$ (in long and short dashed lines respectively) and the unconditional mean exit time $\langle T_r \rangle$ (in solid line) as a function of the resetting position $x_0$. The boundaries are taken at $a=0,~b=3$ for the parameter values: $D=0.5,~r=0.3$. The times coincide for the symmetric case (when $x_0=3/2$),  
as pointed out in the text.}
\label{conditional x0}
\end{figure}

It is also interesting to examine the behavior of the exit times as a function of resetting (initial) position $x_0$. 
Note that, when $x_0=\frac{a+b}{2}$ (i.e., the particle starts from the middle of the interval) we get $\langle t \rangle_{+}=\langle t \rangle_{-}=\frac{2}{r}\sinh\left[ \frac{(b-a)\alpha_0}{4} \right]^2$, and $\langle T_r \rangle=\langle t \rangle_{+}=\langle t \rangle_{-}$ by using \eref{relation-MFT-conditional exit times}. Moreover, if the particle starts from either $x_0=a$ or $x_0=b$, the particle will get absorbed immediately, so that $\langle T_r(a) \rangle=\langle T_r(b) \rangle=0$. However, note that $\langle t(x_0 \to a) \rangle_{+} \neq 0$, which is quite non-intuitive. This is because $\langle t(x_0\to a) \rangle_{+}$ is conditioned on the fact that the particle survives boundary $a$ even if it had started from $x_0 \to a$. 
There will be rare trajectories which will do so, and these will
contribute to $\langle t \rangle_{+}$. A similar argument
justifies the complementary case $\langle t(x_0 \to b) \rangle_{-} \neq 0$. Since these
limits are symmetric with respect to the interval, they are identical and given by $\langle t(x_0 \to a) \rangle_{+} =\langle t(x_0 \to b) \rangle_{-}=\frac{1}{2r} \left[ -1+(b-a)\alpha_0 ~\text{coth}(b-a)\alpha_0  \right]$.
In \fref{conditional x0},
we have plotted the unconditional and the conditional exit times as a function of different $x_0$ . In particular, we have assumed $a=0$ and $b=3$, where we have varied $x_0$. As expected, in the symmetric case, i.e., when $x_0=1.5$, the system does not distinguish between left and right boundaries so that $\langle t \rangle_{+}=\langle t \rangle_{-}=\langle T_r \rangle$. The other limits are also evident from the figure.

\begin{figure}[t]
\centering
\includegraphics[height=6cm, width=8.cm]{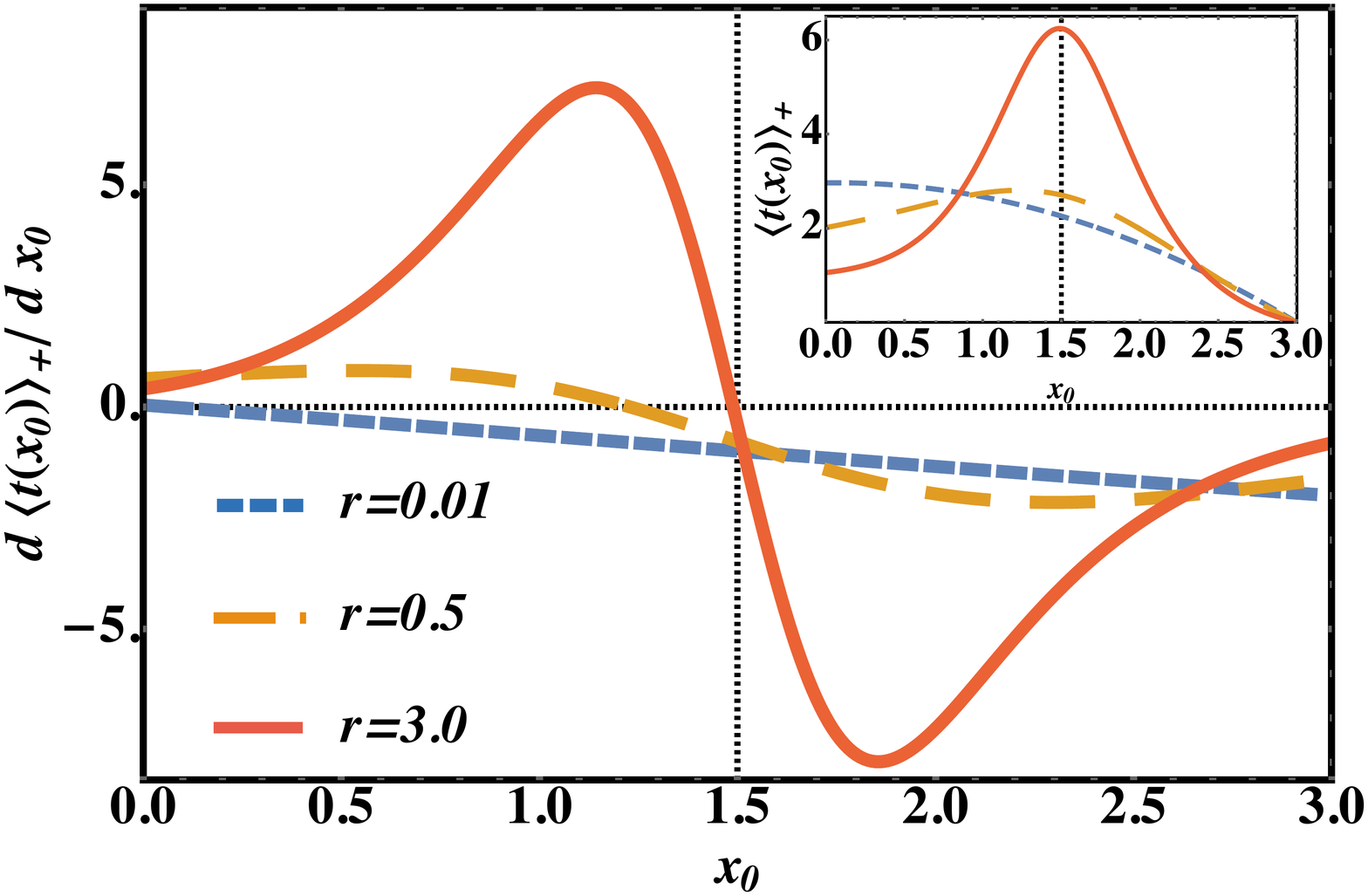}
\caption{(Color online) Plot of the slope $\mathcal{S}(x_0,r)$ as a function of $x_0$ (where $a=0, b=3$) for three different values of restart rate $r=0.01$~(short dashed line), $r=0.5$~(long dashed line), and $r=3.0$~(solid line). The slope $\mathcal{S}(x_0,r)$ becomes zero at $x_0=a$ when $r=0.01$. However, when $r=0.5$, the slope can be both positive or negative with a zero at $x_0$ ($a<x_0<\frac{a+b}{2}$). The behavior remains identical when $r=3.0$ except that the function $\mathcal{S}(x_0,r)$ has a zero at $x_0=\frac{a+b}{2}$, which is the middle of the interval. In the inset we have plotted $\langle t(x_0) \rangle_{+}$ as a function of $x_0$ for the same set of values of $r$ mentioned above. The plot conforms the same behavior as mentioned above.}
\label{slope}
\end{figure}

A closer look at \fref{conditional x0} reveals that both the conditional exit times
$\langle t(x_0) \rangle_{\pm}$ are non-monotonic functions of the initial position $x_0$.
To understand this behavior of $\langle t(x_0) \rangle_{+}$ with respect to $x_0$, we consider the slope
\bea
\mathcal{S}(x_0,r)=\frac{d \langle t(x_0) \rangle_{+}}{dx_0}
\eea
We investigate $\mathcal{S}(x_0,r)$ as a function of $x_0$ by varying $r$ as a parameter. In the $r\to 0$ limit, $\mathcal{S}(x_0,r)=2(a-x_0)$ which is zero only at $x_0=a$, otherwise  negative. Thus $\langle t(x_0) \rangle_{+}$ is strictly monotonic as a function of $x_0$.
In sharp contrast, as $r \to \infty$, we note that $\mathcal{S}(x_0,r)=0$ only when $x_0=\frac{a+b}{2}$. In fact, both $\langle t(x_0) \rangle_{\pm}$ and $\langle T_r \rangle$ will have a single maximum at $x_0=\frac{a+b}{2}$. While
for any finite $r$, the slope $\mathcal{S}(x_0,r)$ becomes zero for $a<x_0<\frac{a+b}{2}$. Consequently, the maximum of the conditional exit time $\langle t(x_0) \rangle_{+}$ spans in the range $a<x_0<\frac{a+b}{2}$. It is worth emphasizing that such non-monotonic behavior of $\langle t(x_0) \rangle_{+}$ is robust to restart, and is lost in the absence of restart when $\langle t(x_0) \rangle_{\pm}$ always become monotonic functions.
We have plotted $\mathcal{S}(x_0,r)$ as a function of $x_0$ for different values of restart rate in \fref{slope}. In the inset we have given plots for $\langle t(x_0) \rangle_{+}$ as a function of $x_0$ for the values of $r$  used in the main plot. A similar analysis also follows
for $\langle t(x_0) \rangle_{-}$ 
which we skip here to avoid redundancy.

\subsection{Restart optimizes success and failure probabilities}
\label{Exit-Time-4}
In the previous sections, we showed how restart can optimize the mean completion time of a first passage process under restart.
In addition, restart can also optimize the success or failure probabilities. This will be the focus of this section. As mentioned before, similar questions were posed and surveyed in a generic set-up in Ref.~\cite{Optimization}. We will refer to Ref.~\cite{Optimization} for the general proofs which was based on the general approach proposed by one of the current authors \cite{PalReuveniPRL}. However, in this paper we have taken different approach and presented alternative proofs of some of the central results.

Let $T_0^{+},T_0^{-}$ be the conditional (success and failure) exit times and  $T_0$ the unconditional exit time (also considered in \sref{MFPT-1}) in the absence of restart.
The density of the latter is given by $f_{T_{0}}(t)=-\frac{dQ_0(x_0,t)}{dt}$, where $Q_0(x_0,t)$ is given by \eref{Q0}. Using this, one can write the Laplace transform $\tilde{T}_0(\lambda) \equiv \int_{0}^{\infty}dt~e^{-\lambda t} f_{T_{0}}(t)$ as
\bea
\tilde{T}_0(\lambda)=2\sum_{n=1}^{\infty}k_n \phi(n)\psi_n(x_0)/\Delta(n,\lambda,0)~.
\eea
The conditional exit time distributions in the absence of restart are identical to the conditional current fluxes through each one of these boundaries [see e.g., \eref{conditional-currents} at the $r\to 0$ limit]. Hence, one can write
\bea
f_{T_0^{+}}(t)&=&J_{+}^{~0}\equiv J_{+}|_{r \to 0}=\frac{2D\pi}{(b-a)^2}\sum_{n=1}^{\infty} n(-1)^{n+1}\psi_n(x_0)e^{-k_n t}~, \nonumber \\
f_{T_0^{-}}(t)&=&J_{-}^{~0} \equiv J_{-}|_{r \to 0}=\frac{2D\pi}{(b-a)^2}\sum_{n=1}^{\infty} n\psi_n(x_0)e^{-k_n t}~.
\label{conditional-FPT-req0}
\eea
One can now compute the Laplace transforms $\tilde{T}_0^{+}(r),\tilde{T}_0^{-}(r)$ evaluated at $r$ using \eref{conditional-FPT-req0}. Furthermore, they can be utilized to recover the success and failure probabilities $\epsilon_{\pm}(x_0)$. These two quantities are related by
\bea
\epsilon_{\pm}(x_0)= \tilde{T}_0^{\pm}(r)/\tilde{T}_0(r)~,
\label{Epsilon}
\eea
and moreover, one can show
\bea
\langle t(x_0) \rangle_{\pm}=\langle T_r(x_0) \rangle-\frac{1}{\epsilon_{\pm}(x_0)}\frac{d \epsilon_{\pm}(x_0)}{dr}~.
\label{tplusminus}
\eea
See \aref{epsilon-proof} and \aref{tplusminus-proof} for a detailed derivation of these two results. It is important to
note that the proofs presented here offer an alternative derivation in comparison to that of demonstrated in Ref.~\cite{Optimization}. 
Remarkably, \eref{tplusminus} offers us a deep insight on how restart could maximize or minimize the chances of our desired outcomes. To see this, let us first consider the case when $\frac{d\epsilon_{+}}{dr}>0$ (see \fref{r-opt} for instance). \eref{tplusminus} then tells us that
$\langle t(x_0) \rangle_{+}<\langle  T_r(x_0) \rangle$ so that the mean conditional time to exit through the boundary $b$ (equivalently rendering `success') is reduced by regulating the restart rate $r$. In other words, by carefully choosing a restart rate, the process can be completed faster in our desired way. On the other hand, when $\frac{d\epsilon_{-}}{dr}<0$, we observe $\langle t(x_0) \rangle_{-}>\langle  T_r(x_0) \rangle$. This implies that the exit time through the boundary $a$  will take longer time, and thus by regulating restart, one can hinder the outcome of `failure'.
The present model at our disposal provides a nice demonstration of how using the restart mechanism one could moderate the chances of desired outcomes of a first passage process that can, in principle, end with multiple eventualities.

\section{Conclusion}
\label{conclusion}
First passage with restart under various set-up has been a focal point in recent studies. The fact that restart has the ability to speedup underlying completion is noteworthy, and this is perhaps the fingerprint of most of these studies. In this paper, we have studied first passage properties of a Brownian particle in a bounded domain in the presence of stochastic resetting. We have shown how this set-up can be visualized as a success-failure problem. We have computed the unconditional mean first passage time and characterized a rich optimization phenomena with respect to the restart rate. We extend the methodology to give estimation of success and failure rates, and furthermore show how restart could optimize the success and failure probabilities. Finally we would like to emphasize that in this problem, we have considered sharp binary outcomes in the sense that the outcomes are either success (full absorption at `$b$') or failure (full absorption at `$a$'). However, these outcomes may not necessarily be sharp, but rather mixed. In other words, the Brownian particle can mix the probabilities of getting absorbed or reflected with certain rate at each boundary \cite{reactive}. We believe that the formalism presented in this paper should be useful to tackle such problems, and is left for future studies.

\section{Acknowledgments} We thank Anupam Kundu for many illuminating discussions. Arnab Pal acknowledges support from the Raymond and Beverly Sackler Post-Doctoral Scholarship.

\appendix

\section{Equivalence between \eref{BFP-qr} and \eref{renewal-qr-LT}}
\label{equivalence}
\noindent
In this section, we will show
the equivalence between \eref{BFP-qr} and \eref{renewal-qr-LT}.
To see this, let us first recall from \eref{renewal-qr-LT} that $q_r(x_0,s)$ contains the following summation $S\equiv \sum_{n=1}^{\infty}~\psi_n(x_0) ~\phi(n)/\Delta(n,r,s)$ both in the numerator and the denominator. To evaluate this sum, we explicitly use the expressions for $\psi_n(x_0)$ and $\phi(n)$, and
rewrite $S$ in the following way
\begin{eqnarray}
S 
&=&\sum_{n=1}^{\infty}~\frac{1}{n\pi}~\frac{\sin(nz)}{\frac{n^2\pi^2D}{(b-a)^2}+r+s}-\sum_{n=1}^{\infty}~\frac{(-1)^n}{n\pi}~\frac{\sin(nz)}{\frac{n^2\pi^2D}{(b-a)^2}+r+s}~,
\label{sum}
\end{eqnarray}
where we have defined $z=\frac{x_0-a}{b-a}\pi$, such that $0<z<\pi$.
We can now make use of the following identities to compute the sums
\begin{eqnarray}
    \sum_{n=1}^{\infty}~\frac{1}{n}~\frac{\sin(nz)}{n^2+\beta^2}&=& \frac{\pi}{2\beta^2}~\frac{\cosh(\beta \pi)\sinh(\beta z)-\cosh(\beta z)\sinh(\beta \pi)}{\sinh(\beta\pi)} \nonumber \\ &-&\frac{z}{2\beta^2}+\frac{\pi}{2\beta^2},~~~~~~~0\leq z\leq 2\pi 
    \end{eqnarray}
and 
\bea
    \sum_{n=1}^{\infty}~\frac{(-1)^n}{n}~\frac{\sin(nz)}{n^2+\beta^2}&=&
        \frac{\pi}{2\beta^2}~\frac{\sinh(\beta z)}{\sinh(\beta \pi)}-\frac{z}{2\beta^2},-\pi\leq z\leq \pi. \nonumber \\
\eea
Using the identities above, we can now convert the sums in \eref{sum}. After doing a bit of simplification we arrive at
\begin{equation}
S=\frac{1-g_r(x_0,s)}{2(r+s)} \nonumber~,
\end{equation}
where $g_r(x_0,s)$ is given by \eref{gr}
\bea
g_r(x_0,s)=\frac{\sinh(b-x_0)\alpha+\sinh(x_0-a)\alpha}{\sinh(b-a)\alpha}~.
\eea
Substituting $S$ in  \eref{renewal-qr-LT}, we obtain
\begin{equation}
q_r(x_0,s)=\frac{2S}{1-2rS}=\frac{1-g_r(x_0,s)}{s+rg_r(x_0,s)} \nonumber~,
\end{equation}
which is indeed \eref{BFP-qr}, as reported earlier.

\section{Proof of \eref{Epsilon}}
\label{epsilon-proof}
In this section, we provide a proof for the following relations
\bea
\epsilon_{\pm}(x_0)= \tilde{T}_0^{\pm}(r)/\tilde{T}_0(r)
\eea
Here, we will demonstrate the proof for
$\epsilon_{+}(x_0)$ while an analysis for $\epsilon_{-}(x_0)$ can also be made in a similar manner. Let us first recall the definition from \eref{epsilon}
\bea
\epsilon_{+}(x_0)=\int_0^\infty~dt~J_{+}(x_0,t)=j_{+}(x_0,s=0)~,
\eea
where $j_{+}(x_0,s)$ follows from \eref{conditional-currents-LT}
\bea
j_{+}(x_0,s)&=&-D\frac{\partial p_r(x,s|x_0,0)}{\partial x}|_{x=b}~.
\eea
which can be rewritten as
\bea
j_{+}(x_0,s)&=&\frac{1}{1-r q_0(x_0,s+r)}~\left[ -D\frac{\partial p_0(x,s+r|x_0,0)}{\partial x}|_{x=b}\right]~,\nonumber \\
\eea
by replacing $p_r(x,s|x_0,0)$ with $p_0(x,s|x_0,0)$ using \eref{propagator-formula}. Setting $s=0$ in the above equation, we find
\bea
\epsilon_{+}(x_0)&=&j_{+}(x_0,s=0)~ \nonumber\\
&=&\frac{1}{1-r q_0(x_0,r)}~\left[ -D\frac{\partial p_0(x,r|x_0,0)}{\partial x}|_{x=b}\right]~.
\label{epsilonseq0}
\eea
We now note that 
\bea
\tilde{T}_0(r)&=&\int_0^\infty~dt~e^{-rt}~f_{T_0}(t) \nonumber\\
&=&\int_0^\infty~dt~e^{-rt}~\left[ -\frac{dQ_0}{dt} \right]
\label{t0}
\eea
where $Q_0(x_0,t)$ is the survival probability (see the results from \sref{Exit-Time-4}). Doing an integration by parts in \eref{t0}, we find
\bea
\tilde{T}_0(r)=1-rq_0(x_0,r)~.
\label{R1}
\eea
On the other hand, let us define
\bea
j_{+}^{~0}(x_0,r)& \equiv &\int_0^\infty~dt~e^{-rt} ~J_{+}^{~0}(x_0,t) \nonumber \\
&=&\int_0^\infty~dt~e^{-rt} ~\left[ -D\frac{\partial P_0(x,t|x_0,0)}{\partial x}|_{x=b}\right] \nonumber\\
&=& \left[ -D\frac{\partial p_0(x,r|x_0,0)}{\partial x}|_{x=b}\right]
\eea
where $J_{+}^{~0}(x_0,t)$ was introduced in \eref{conditional-FPT-req0}. But also note that $J_{+}^{~0}(x_0,t)=f_{T_0^{+}}(t)$. Hence
\bea
\tilde{T}_{0}^{+}(r) &=& \int_0^\infty~dt~e^{-rt}  ~f_{T_0^{+}}(t) \nonumber \\
&=&\int_0^\infty~dt~e^{-rt} ~J_{+}^{~0}(x_0,t) \nonumber \\
&=&j_{+}^{~0}(x_0,r).
\label{lt-current}
\eea
Thus, we have proven
\bea
j_{+}^{~0}(x_0,r)=\tilde{T}_{0}^{+}(r)=\left[ -D\frac{\partial p_0(x,r|x_0,0)}{\partial x}|_{x=b}\right].
\label{R2}
\eea
Using \eref{R1} and \eref{R2} in \eref{epsilonseq0}, we finally arrive at the following relation we were seeking after
\bea
\epsilon_{+}(x_0)=\frac{\tilde{T}_{0}^{+}(r)}{\tilde{T}_0(r)}~.
\eea

\section{Proof of \eref{tplusminus}}
\label{tplusminus-proof}
In this Appendix, we will provide the proofs of the following relations
\bea
\langle t(x_0) \rangle_{\pm}=\langle T_r(x_0) \rangle-\frac{1}{\epsilon_{\pm}(x_0)}\frac{d \epsilon_{\pm}(x_0)}{dr}~.
\eea
Like in the preceding Appendix, here we will only demonstrate the proof for $\langle t(x_0) \rangle_{+}$ and leave $\langle t(x_0) \rangle_{-}$ for a likewise proof. 
\noindent
We start by writing a renewal equation for the current $J_{+}(x_0,t)$ given by
\bea
J_{+}(x_0,t)=e^{-rt}~J_{+}^{~0}(x_0,t)  +r \int_0^t~d\tau~e^{-r\tau} ~J_{+}^{~0}(x_0,\tau)~Q_r(x_0,t-\tau).\nonumber \\
\eea
Such kind of renewal equations served as a bedrock already for many of our central results.  
Taking Laplace transform on both sides of the above equation, we obtain
\bea
j_{+}(x_0,s)&=&j_{+}^{~0}(x_0,s+r) \left[ 1+r q_r(x_0,s) \right] \nonumber \\
&=&\frac{j_{+}^{~0}(x_0,s+r)}{1-rq_0(x_0,s+r)}~,
\label{j+}
\eea
where we have used \eref{renewal-qr}. Now, recall from \eref{mean-exit-times-definition-LT} 
\bea
\langle t(x_0) \rangle_{+}=\frac{-\frac{\partial j_{+}(x_0,s)}{\partial s}|_{s \to 0}}{j_{+}(x_0,s=0)}~,
\eea
which can be rewritten as 
\bea
\langle t(x_0) \rangle_{+}=- \left[ \frac{\partial}{\partial s} \ln j_{+}(x_0,s) \right]_{s\to 0}
\eea
Substituting $j_{+}(x_0,s)$ from \eref{j+} in the above equation gives us
\bea
\langle t(x_0) \rangle_{+}= - \left[ \frac{\partial}{\partial s} \ln j_{+}^{~0}(x_0,s+r) \right]_{s\to 0}
+\left[ \frac{\partial}{\partial s} \ln \left(1-r q_0(x_0,s+r) \right) \right]_{s\to 0} \nonumber \\
 \label{t-fracs}
\eea
We have already proven in the last section that
$\tilde{T}_{0}^{+}(s)=j_{+}^{~0}(x_0,s)$ [see e.g., \eref{lt-current}], such that $j_{+}^{~0}(x_0,s+r)|_{s \to 0}=\tilde{T}_{0}^{+}(r)$. Furthermore, it is also evident from \eref{lt-current} that
\bea
\frac{\partial j_{+}^{~0}(x_0,s+r)}{\partial s}\bigg|_{s \to 0} = \frac{d}{dr} \tilde{T}_{0}^{+}(r)
\label{fracs2}
\eea
On the other hand, note that
\bea
&~&\left[ \frac{\partial}{\partial s} \ln \left(1-r q_0(x_0,s+r) \right) \right]_{s\to 0}\nonumber \\
&=&\frac{1}{1-rq_0(x_0,s+r)}\bigg|_{s\to 0}~
\times~\frac{\partial}{\partial s} \left[ 1-rq_0(x_0,s+r) \right]\bigg|_{s\to 0} \nonumber \\
&=&\frac{1}{1-rq_0(x_0,s+r)}\bigg|_{s\to 0}~
\times~ \left[ -r \frac{\partial q_0(x_0,s+r)}{\partial s} \right]_{s\to 0} \nonumber \\
&=&\frac{r}{1-rq_0(x_0,r)}~\int_0^\infty~dt~t~e^{-rt}~Q_0(x_0,t) \nonumber \\
&=&\frac{r}{1-rq_0(x_0,r)}~\left[ -\frac{d q_0(x_0,r)}{dr} \right]
\label{fracs}
\eea
\\
Now from \eref{R1}, recall that $\tilde{T}_0(r)=1-rq_0(x_0,r)$. Taking derivative on both sides with respect to $r$, we find
\bea
\frac{d \tilde{T}_0(r)}{dr}=-r~\frac{d q_0(x_0,r)}{dr}-q_0(x_0,r)
\eea
\\
Substituting $d q_0(x_0,r)/dr$ from the above expression in \eref{fracs}, we arrive at
\bea
&~&\left[ \frac{\partial}{\partial s} \ln \left(1-r q_0(x_0,s+r) \right) \right]_{s\to 0}=\frac{1}{\tilde{T}_0(r)} \left[ q_0(x_0,r)+ \frac{d \tilde{T}_0(r)}{dr} \right]~. \nonumber \\
\label{fracs1}
\eea
\\
Replacing all the expressions obtained from \eref{fracs2} and \eref{fracs1} in \eref{t-fracs}, we finally arrive at
\bea
\langle t(x_0) \rangle_{+}&=&-\frac{1}{\tilde{T}_0^{+}(r)}\frac{d \tilde{T}_0^{+}(r)}{dr}+\frac{1}{\tilde{T}_0(r)} \left[ q_0(x_0,r)+ \frac{d \tilde{T}_0(r)}{dr} \right] \nonumber \\
&=&\frac{q_0(x_0,r)}{\tilde{T}_0(r)}-\frac{1}{\tilde{T}_0^{+}(r)}\frac{d \tilde{T}_0^{+}(r)}{dr}+\frac{1}{\tilde{T}_0(r)}\frac{d \tilde{T}_0(r)}{dr}\nonumber \\
&=&\frac{1-\tilde{T}_0(r)}{r\tilde{T}_0(r)}-\frac{d}{dr} \ln 
\tilde{T}_0^{+}(r) + \frac{d}{dr} \ln \tilde{T}_0(r)
\nonumber \\
&=& \langle  T_r \rangle -\frac{d}{dr} \ln \frac{\tilde{T}_0^{+}(r)}{\tilde{T}_0(r)} \nonumber \\
&=& \langle  T_r \rangle -\frac{d}{dr} \ln \epsilon_{+}(x_0) \nonumber \\
&=& \langle  T_r \rangle -\frac{1}{\epsilon_{+}(x_0)}\frac{d\epsilon_{+}(x_0)}{dr}~ 
\eea
which is our desired result \eref{tplusminus}. This completes the proof.


\begin{thebibliography}{1}

\bibitem{Restart1} Evans, M.R. and Majumdar, S.N., 2011. Diffusion with stochastic resetting. Physical review letters, 106(16), p.160601.

\bibitem{Restart2} Evans, M.R. and Majumdar, S.N., 2011. Diffusion with optimal resetting. Journal of Physics A: Mathematical and Theoretical, 44(43), p.435001.

\bibitem{KM}
Evans, M.R., Majumdar, S.N. and Mallick, K., 2013. Optimal diffusive search: nonequilibrium resetting versus equilibrium dynamics. Journal of Physics A: Mathematical and Theoretical, 46(18), p.185001.

\bibitem{Restart3} Montero, M. and Villarroel, J., 2013.
Monotonic continuous-time random walks with drift and stochastic reset
events. Physical Review E, 87(1), p.012116.

\bibitem{restart_conc1} Gupta, S., Majumdar, S.N. and Schehr, G., 2014. Fluctuating interfaces subject to stochastic resetting. Physical review letters, 112(22), p.220601.

\bibitem{restart_conc2} Pal, A., 2015. Diffusion in a potential landscape with stochastic resetting. Physical Review E, 91(1), p.012113.

\bibitem{restart_conc3} Eule, S. and Metzger, J.J., 2016. Non-equilibrium steady states of stochastic processes with intermittent resetting. New Journal of Physics, 18(3), p.033006.

\bibitem{restart_conc4} Durang, X., Henkel, M. and Park, H., 2014. The statistical mechanics of the coagulation–diffusion process with a stochastic reset. Journal of Physics A: Mathematical and Theoretical, 47(4), p.045002.

\bibitem{restart_conc5} Majumdar, S.N., Sabhapandit, S. and Schehr, G., 2015. Dynamical transition in the temporal relaxation of stochastic processes under resetting. Physical Review E, 91(5), p.052131.

\bibitem{restart_conc6} Evans, M.R. and Majumdar, S.N., 2014. Diffusion with resetting in arbitrary spatial dimension. Journal of Physics A: Mathematical and Theoretical, 47(28), p.285001.

\bibitem{restart_conc7} M\'endez, V. and Campos, D., 2016. Characterization of stationary states in random walks with stochastic resetting. Physical Review E, 93(2), p.022106.

\bibitem{restart_conc8} Christou, C. and Schadschneider, A., 2015. Diffusion with resetting in bounded domains. Journal of Physics A: Mathematical and Theoretical, 48(28), p.285003.

\bibitem{restart_conc9} Chatterjee, A., Christou, C. and Schadschneider, A., 2018. Diffusion with resetting inside a circle. Physical Review E, 97(6), p.062106.

\bibitem{restart_conc11} Falc\'on-Cort\'es, A., Boyer, D., Giuggioli, L. and Majumdar, S.N., 2017. Localization transition induced by learning in random searches. Physical review letters, 119(14), p.140603.

\bibitem{restart_conc12} Falcao, R. and Evans, M.R., 2017. Interacting Brownian motion with resetting. Journal of Statistical Mechanics: Theory and Experiment, 2017(2), p.023204.

\bibitem{restart_conc15} Montero, M., Mas\'o-Puigdellosas, A. and Villarroel, J., 2017. Continuous-time random walks with reset events: Historical background and new perspectives.  J. Eur. Phys. J. B 90: 176.

\bibitem{restart_conc16} Majumdar, S.N., Sabhapandit, S. and Schehr, G., 2015. Random walk with random resetting to the maximum position. Physical Review E, 92(5), p.052126.

\bibitem{restart_conc17} Meylahn, J.M., Sabhapandit, S. and Touchette, H., 2015. Large deviations for Markov processes with resetting. Physical Review E, 92(6), p.062148.

\bibitem{restart_conc18} Boyer, D., Evans, M.R. and Majumdar, S.N., 2017. Long time scaling behaviour for diffusion with resetting and memory. Journal of Statistical Mechanics: Theory and Experiment, 2017(2), p.023208.


\bibitem{restart_conc21} Shkilev, V.P., 2017. Continuous-time random walk under time-dependent resetting. Physical Review E, 96(1), p.012126.


\bibitem{tethered}
Giuggioli, L., Gupta, S. and Chase, M., 2018. Comparison of two models of tethered motion, J. Phys. A  52, 075001 (2019). 

\bibitem{localtimer}
Pal, A., Chatterjee, R., Reuveni, S. and Kundu, A., 2019. Local time of diffusion with stochastic resetting, arXiv preprint arXiv:1902.00907.


\bibitem{Satya-refractory}
Evans, M.R. and Majumdar, S.N.,  Effects of refractory period on stochastic resetting, J. Phys. A  52, 01LT01 (2018).

\bibitem{Satya-RT}
Evans, M.R. and Majumdar, S.N., 2018. Run and tumble particle under resetting: a renewal approach. Journal of Physics A: Mathematical and Theoretical, 51(47), p.475003.


\bibitem{Restart4}Pal, A., Kundu, A. and Evans, M.R., 2016.
Diffusion under time-dependent resetting. Journal of Physics A: Mathematical
and Theoretical, 49(22), p.225001.

\bibitem{Restart5}Nagar, A. and Gupta, S., 2016. Diffusion
with stochastic resetting at power-law times. Physical Review E, 93(6),
p.060102.

\bibitem{transport1}Mas\'o-Puigdellosas, A., Campos, D. and M\'endez, V., 2018. Stochastic processes subject to a reset-and-residence mechanism: transport properties and first arrival statistics, J. Stat. Mech.~(2019) 033201.

\bibitem{transport2} Mas\'o-Puigdellosas, A., Campos, D. and M\'endez, V., 2018. Transport properties and first arrival statistics of random searches with stochastic reset times,  Phys. Rev. E 99, 012141 (2019).

\bibitem{Restart-Search1}Kusmierz, L., Majumdar, S.N., Sabhapandit, S. and Schehr, G., 2014. First order transition for the optimal search time of L\'evy flights with resetting. Physical review letters, 113(22), p.220602.

\bibitem{Restart-Search2}Kusmierz, L. and Gudowska-Nowak,
E., 2015. Optimal first-arrival times in L\'evy flights with resetting.
Physical Review E, 92(5), p.052127.

\bibitem{Restart-Search3}Bhat, U., De Bacco, C. and Redner, S., 2016. Stochastic search with Poisson and deterministic resetting. Journal of Statistical Mechanics: Theory and Experiment, 2016(8), p.083401.

\bibitem{Chechkin}Chechkin, A. and Sokolov, I.M., 2018. Random search with resetting: a unified renewal approach. Physical review letters, 121(5), p.050601.

\bibitem{restart_conc19} Husain, K. and Krishna, S., 2016. Efficiency of a Stochastic Search with Punctual and Costly Restarts. arXiv preprint arXiv:1609.03754.

\bibitem{restart_conc20} Campos, D. and M\'endez, V., 2015. Phase transitions in optimal search times: How random walkers should combine resetting and flight scales. Physical Review E, 92(6), p.062115.


\bibitem{ReuveniPRL}Reuveni, S., 2016. Optimal stochastic
restart renders fluctuations in first passage times universal. Physical
review letters, 116(17), p.170601.

\bibitem{PalReuveniPRL} Pal, A. and Reuveni, S., 2017. First Passage under Restart. Physical review letters, 118(3), p.030603.


\bibitem{Optimization} Belan, S., 2018. Restart could optimize the probability of success in a Bernoulli trial. Physical review letters, 120(8), p.080601.

\bibitem{drift-diffusion}
Ray, S., Mondal, D. and Reuveni, S., 2018. P\'eclet number governs transition to acceleratory restart in drift-diffusion. arXiv preprint arXiv:1811.08239.

\bibitem{branching}
Pal, A., Eliazar, I. and Reuveni, S., 2019. First passage under restart with branching. Physical review letters, 122(2), p.020602.

\bibitem{RednerBook}Redner, S., 2007. A Guide to First-Passage Processes. A Guide to First-Passage Processes, by Sidney Redner, Cambridge, UK: Cambridge University Press, 2007.

\bibitem{MetzlerBook}Metzler, R., Redner, S. and Oshanin, G.,
2014. First-Passage Phenomena and Their Applications (Vol. 35). Singapore: World Scientific.

\bibitem{Schehr-review}Bray, A.J., Majumdar, S.N. and Schehr, G., 2013. Persistence and first-passage properties in nonequilibrium systems. Advances in Physics, 62(3), pp.225-361.

\bibitem{Benichou-review} B\'enichou, O., Loverdo, C., Moreau, M. and Voituriez, R., 2011. Intermittent search strategies. Reviews of Modern Physics, 83(1), p.81.


\bibitem{Restart-Biophysics1}Reuveni, S., Urbakh, M. and
Klafter, J., 2014. Role of substrate unbinding in Michaelis-Menten
enzymatic reactions. Proceedings of the National Academy of Sciences,
111(12), pp.4391-4396.


\bibitem{restart-CS1} Luby, M., Sinclair, A. and Zuckerman, D., 1993, June. Optimal speedup of Las Vegas algorithms. In Theory and Computing Systems, 1993., Proceedings of the 2nd Israel Symposium on the (pp. 128-133). IEEE.


\bibitem{Restart-Biophysics2}Rotbart, T., Reuveni, S. and Urbakh, M., 2015. Michaelis-Menten reaction scheme as a unified approach towards the optimal restart problem. Physical Review E, 92(6), p.060101.

\bibitem{Restart-Biophysics3}Robin, T., Reuveni, S. and Urbakh, M., 2018. Single-molecule theory of enzymatic inhibition. Nature communications, 9(1), p.779.

\bibitem{Restart-Biophysics4} Rold\'an, \'E., Lisica, A., S\'anchez-Taltavull, D. and Grill, S.W., 2016. Stochastic resetting in backtrack recovery by RNA polymerases. Physical Review E, 93(6), p.062411.

\bibitem{restart_thermo1} Fuchs, J., Goldt, S. and Seifert, U., 2016. Stochastic thermodynamics of resetting. EPL (Europhysics Letters), 113(6), p.60009.

\bibitem{restart_thermo2} Pal, A. and Rahav, S., 2017. Integral fluctuation theorems for stochastic resetting systems. Physical Review E, 96(6), p.062135.

\bibitem{Quantum1} Rose, D.C., Touchette, H., Lesanovsky, I. and Garrahan, J.P., 2018. Spectral properties of simple classical and quantum reset processes. Physical Review E 98, 022129.

\bibitem{Quantum2} Mukherjee, B., Sengupta, K. and Majumdar, S.N., 2018. Quantum dynamics with stochastic reset. Physical Review B 98, 104309.

\bibitem{reactive}
Pal, A., Castillo, I.P. and Kundu, A., 2018. Motion of a Brownian molecule in the presence of reactive boundaries. arXiv preprint arXiv:1805.04762.

\end{thebibliography}
\end{document}